\documentclass{article}
\usepackage{amsmath,amssymb,enumerate,color,framed}

\newcommand{\assign}{:=}
\newcommand{\nocomma}{}
\newcommand{\tmem}[1]{{\em #1\/}}
\newcommand{\tmmathbf}[1]{\ensuremath{\boldsymbol{#1}}}
\newcommand{\tmop}[1]{\ensuremath{\operatorname{#1}}}
\newcommand{\tmstrong}[1]{\textbf{#1}}
\newcommand{\tmtextbf}[1]{{\bfseries{#1}}}
\newcommand{\tmtextit}[1]{{\itshape{#1}}}
\newenvironment{enumerateroman}{\begin{enumerate}[i.] }{\end{enumerate}}
\newenvironment{proof}{\noindent\textbf{Proof\ }}{\hspace*{\fill}$\Box$\medskip}
\newenvironment{tmindent}{\begin{tmparmod}{1.5em}{0pt}{0pt} }{\end{tmparmod}}
\newenvironment{tmparmod}[3]{\begin{list}{}{\setlength{\topsep}{0pt}\setlength{\leftmargin}{#1}\setlength{\rightmargin}{#2}\setlength{\parindent}{#3}\setlength{\listparindent}{\parindent}\setlength{\itemindent}{\parindent}\setlength{\parsep}{\parskip}} \item[]}{\end{list}}
\newenvironment{tmparsep}[1]{\begingroup\setlength{\parskip}{#1}}{\endgroup}
\definecolor{grey}{rgb}{0.75,0.75,0.75}
\definecolor{orange}{rgb}{1.0,0.5,0.5}
\definecolor{brown}{rgb}{0.5,0.25,0.0}
\definecolor{pink}{rgb}{1.0,0.5,0.5}

\begin{document}

\title{Directed Max Flow}\author{Cheng Wang}\maketitle

Let $G = \left( V, E \right)$ be an directed graph with n vertices and m
edges, a {\tmem{source}} vertex s and a {\tmem{sink}} vertex t, capacity $u_e$
for each edge $e$. My method to find the approximate directed max flow of $G$
have two steps. The first step is to reduce it to an undirected problem, then
the second step is to solve the undirected problem via a variation of the
multiplicative weights update method in the marvelous paper [E].

Through this article, we fix a $\varepsilon$ $\left( 0 < \varepsilon < 1 / 2
\right)$.

\section{Reduction}

We construct the related undirected graph $\tilde{G} = \left( \tilde{V}
\nocomma, \tilde{E} \right)$ as follows(which is well-known in graph theory).
We take $\tilde{V}$ to be equal $V$. Next, for each arc $e = \left( u, v
\right)$ of G we add (undirected) edges e, $e_s : = \left( s, v \right)$, and
$e_t \assign \left( u, t \right)$ to $\tilde{G}$, with the capacity of $e$
being $u_e$, the capacity of $e_s, e_t$ being $\left( 1 + \varepsilon \right)
u_e$ (i.e. {\color{red} $u_{e_s} = u_{e_t} = \left( 1 + \varepsilon \right)
u_e$}). We allow $\tilde{G}$ to have multi-edges. By the max-flow min-cut
theorem, if the max flow value of $G$ is $F^{\ast}$, then the max flow of
$\tilde{G}$ is $\left( 2 + \varepsilon \right) F^{\ast} + \sum_{e \in E}
\left( 1 + \varepsilon \right) u_e$.

We will use the following definition:

{\noindent}\tmtextbf{Definition. }\tmtextit{({\tmstrong{Magic Solver}}). For
    $\varepsilon > 0$, a Magic Solver for $\tilde{G}$ is an algorithm that, given a real number $F
    > \sum_{e \in E} \left(1+\varepsilon\right)u_e$, works as follows:
\begin{enumerate}
  \item If $F \leqslant \left( 2 + \varepsilon \right) F^{\ast} + \sum_{e \in
  E} \left( 1 + \varepsilon \right) u_e$, then it outputs an s-t flow f of
  $\tilde{G}$ satisfying:
  \begin{enumerateroman}
    \item $\left| f \right| = F$
    
    \item $- \left( 1 + \varepsilon \right) u_e \leqslant f \left( e \right)
    \leqslant \left( 1 + \varepsilon \right) u_e$, for all $e \in \tilde{G}$
  \end{enumerateroman}
  \item If $F > \left( 2 + \varepsilon \right) F^{\ast} + \sum_{e \in E}
  \left( 1 + \varepsilon \right) u_e$, then it either outputs a flow f
  satisfying conditions (i), (ii) or outputs ``{\tmstrong{fail}}''.
\end{enumerate}}

In fact, our directed max flow problem can be reduced to implement such
{\tmem{Magic Solver}}.

{\noindent}\tmtextbf{Lemma. }\tmtextit{Given a Magic Solver for $\tilde{G}$,
we can get a $\left( 1 - \varepsilon \right)$-approximation of $F^{\ast}$ for
$G$ using binary search.}{\hspace*{\fill}}{\medskip}

\begin{proof}
  Given any $F > 0$.
  
  If $F \leqslant F^{\ast}$, we apply the {\tmem{Magic Solver}} to $2 F +
  \sum_{e \in E} \left( 1 + \varepsilon \right) u_e$, it will return a flow
  $\tilde{f}$ of $\tilde{G}$ satisfying (ii). For each arc $e \in E$, let
  $f^e$ be the s-t flow send $\left( 1 + \varepsilon \right) u_e$ units of
  flow through $e_s$, e, $e_t$. Let $\tilde{f}'$ be a flow out of $\tilde{f}$
  by substracting for each $e \in E$, the s-t flow $f^e$, i.e.
  \[ \tilde{f}' = \frac{1}{2} \left( \tilde{f} - \sum_{e \in E} f^e \right) .
  \]
  So for all $e \in E$, we have $0 \leqslant \tilde{f}' \left( e \right) \leqslant \left( 1 + \varepsilon \right) u_e$, $-\left( 1 + \varepsilon \right) u_e \leqslant \tilde{f}'(e_s) \leqslant 0$, $-\left( 1 + \varepsilon \right) u_e \leqslant \tilde{f}'(e_s) \leqslant 0$ and $\left| \tilde{f}' \right| = F$.
  That is to say the only
  direction the flow $\tilde{f}'$ over $e_s$ (resp. $e_t$) is towards $s$
  (resp. out of $t$). If we apply flow-cycle-canceling algorithm to
  $\tilde{f}'$, then the acyclic flow $f$ we obtain will be directed flow of
  $G$ satisfying $0 \leqslant f \left( e \right) \leqslant \left( 1 +
  \varepsilon \right) u_e$ for all $e \in E$. So if $F \leqslant F^{\ast}$, we
  can obtain a feasible flow of $G$ at least $F / \left( 1 + \varepsilon
  \right)$.
  
  If $F > F^{\ast}$, Same as the above analysis, we will either obtain a
  feasible flow of G at least $F / \left( 1 + \varepsilon \right)$ or get a
  ``{\tmstrong{fail}}''.
  
  All these things allow us to find a $\left( 1 - \varepsilon
  \right)$-approximation of $F^{\ast}$ using binary search.
\end{proof}

\section{Magic Solver}

We will show that we can implement a {\tmstrong{{\tmem{Magic Solver}}}} for
$\tilde{G}$ via the multiplicative weights update method in [E].

First, we introduce some notations for $\tilde{G}$. If $\left\{ \omega_e
\right\}_{e \in E}$ be the weights of $G$, then we let $\omega_{e_s} = w_{e_t}
= w_e$ in $\tilde{G}$. Also we use these notations $\left| w \right|_1 \assign \sum_{e \in E}
w_e$ and $\left| \tilde{w} \right|_1 \assign \sum_{e \in \tilde{E}} w_e$.
Apparently, we have $\left| \tilde{w} \right|_1 = 3 \left| w \right|_1$. We
define the {\tmem{congestion}}, for $e \in E$, as follows,
\[ {\color{red} \tmmathbf{c}\tmmathbf{o}\tmmathbf{n}\tmmathbf{g}_f \left( e
   \right) =\tmmathbf{c}\tmmathbf{o}\tmmathbf{n}\tmmathbf{g}_f \left( e_s
   \right) =\tmmathbf{c}\tmmathbf{o}\tmmathbf{n}\tmmathbf{g}_f \left( e_t
   \right) \assign \left| \frac{f \left( e \right)}{u_e} \right|} . \]

If we can design an $\left( \varepsilon, \rho \right)$ oracle with respect to
our new definition of {\tmem{congestion}} via the following algorithm, then by
the multiplicative-weights-update routine in [E] we can implement a
{\tmstrong{{\tmem{Magic Solver}}}}.

\begin{framed}
  {\noindent}\begin{tmparsep}{0em}
    \tmtextbf{Algorithm}{\smallskip}
    
    \begin{tmindent}
      {\tmstrong{Input}}: The graph $\tilde{G}$, with capacities $\left\{ u_e
      \right\}_{e \in \tilde{E}}$, a target flow value F, and edge weight
      $\left\{ w_e \right\}_{e \in E}$
      
      {\tmstrong{Output}}: Either a flow $\tilde{f}$, or ``{\tmstrong{fail}}''
      indicating that F
      
      for each $e \in E$, $r_e = r_{e_s} = r_{e_t} \leftarrow \frac{1}{u_e^2}
      \left( \omega_e + \frac{\varepsilon \left| w \right|_1}{3 m} \right)$
      
      Find an $\left( {\color{red} \frac{\varepsilon}{10}}
      \right)$-approximate electrical flow $\tilde{f}$ using Theorem 2.3 in
      [E] on $\tilde{G}$ with resistances {\tmstrong{$r$}} and target flow
      value F
      
      {\tmstrong{If}} $\mathcal{E}_r \left( \tilde{f} \right) > \left( 1 + \frac{\varepsilon}{10} \right)\left( 1 + \frac{\varepsilon}{3} \right)\frac{1+2\left(1 + \varepsilon\right)^2}{3}\left| \tilde{w} \right|_1$
      {\tmstrong{then return ``fail''}}
      
      {\tmstrong{else return}} $\tilde{f}$
    \end{tmindent}
  \end{tmparsep}{\hspace*{\fill}}{\medskip}
\end{framed}

We only need to show that the above algorithm implement an $\left(
\varepsilon, \rho \right)$-oracle.

Suppose $f^{\ast}$ is a maximum flow of $\tilde{G}$. By its feasibility, for
all $e \in E$ $\tmmathbf{c}\tmmathbf{o}\tmmathbf{n}\tmmathbf{g}_{f^{\ast}}
\left( e \right) \leqslant 1$,
$\tmmathbf{c}\tmmathbf{o}\tmmathbf{n}\tmmathbf{g}_{f^{\ast}} \left( e_s
\right) \leqslant \left( 1 + \varepsilon \right)$ and
$\tmmathbf{c}\tmmathbf{o}\tmmathbf{n}\tmmathbf{g}_{f^{\ast}} \left( e_t
\right) \leqslant \left( 1 + \varepsilon \right)$, so
\begin{eqnarray*}
  \mathcal{E}_r \left( f^{\ast} \right) & = & \sum_{e \in \tilde{E}} \left(
  w_e + \frac{\varepsilon \left| w \right|_1}{3 m} \right) \left(
  \tmmathbf{c}\tmmathbf{o}\tmmathbf{n}\tmmathbf{g}_{f^{\ast}} \left( e \right)
  \right)^2\\
  & \leqslant & \left( \sum_{e \in E} \left( w_e + \frac{\varepsilon \left| w
  \right|_1}{3 m} \right) \right) \left( 1 + 2 \left( 1 + \varepsilon
  \right)^2 \right)\\
  & = & \left( 1 + \frac{\varepsilon}{3} \right) \frac{1 + 2 \left( 1 +
  \varepsilon \right)^2}{3} \left| \tilde{w} \right|_1 .
\end{eqnarray*}
This implies
\[ \mathcal{E}_r \left( \tilde{f} \right) \leqslant \left( 1 +
   \frac{\varepsilon}{10} \right) \mathcal{E}_r \left( f^{\ast} \right)
   \leqslant \left( 1 + \frac{\varepsilon}{10} \right) \left( 1 +
   \frac{\varepsilon}{3} \right) \frac{1 + 2 \left( 1 + \varepsilon
   \right)^2}{3} \left| \tilde{w} \right|_1 . \]
This implies
\begin{equation}
  \sum_{e \in \tilde{E}} w_e \left(
  \tmmathbf{c}\tmmathbf{o}\tmmathbf{n}\tmmathbf{g}_{\tilde{f}} \left( e
  \right) \right)^2 \leqslant \left( 1 + \frac{\varepsilon}{10} \right) \left(
  1 + \frac{\varepsilon}{3} \right) \frac{1 + 2 \left( 1 + \varepsilon
  \right)^2}{3} \left| \tilde{w} \right|_1,
\end{equation}
and for all $e \in \tilde{E}$,
\begin{equation}
  \frac{\varepsilon \left| w \right|_1}{3 m} \left(
  \tmmathbf{c}\tmmathbf{o}\tmmathbf{n}\tmmathbf{g}_{\tilde{f}} \left( e
  \right) \right)^2 \leqslant \left( 1 + \frac{\varepsilon}{10} \right) \left(
  1 + \frac{\varepsilon}{3} \right) \frac{1 + 2 \left( 1 + \varepsilon
  \right)^2}{3} \left| \tilde{w} \right|_1 \leqslant 3 \left| \tilde{w}
  \right|_1 .
\end{equation}
So by Equation (1) and Cauchy-Schwartz inequality,
\begin{equation}
  \sum_{e \in \tilde{E}} w_e
  \tmmathbf{c}\tmmathbf{o}\tmmathbf{n}\tmmathbf{g}_{\tilde{f}} \left( e
  \right) \leqslant \sqrt{\left( 1 + \frac{\varepsilon}{10} \right) \left( 1 +
  \frac{\varepsilon}{3} \right) \frac{1 + 2 \left( 1 + \varepsilon
  \right)^2}{3}} \left| \tilde{w} \right|_1 < \left( 1 + \varepsilon \right)
  \left| \tilde{w} \right|_1 .
\end{equation}
And Equation (2) implies that
\begin{equation}
  \tmmathbf{c}\tmmathbf{o}\tmmathbf{n}\tmmathbf{g}_{\tilde{f}} \left( e
  \right) \leqslant \sqrt{27 m / \varepsilon} .
\end{equation}
So our algorithm implements an $\left( \varepsilon \nocomma \nocomma, \sqrt{27
m / \varepsilon} \right)$-oracle, which implies we have designed a
{\tmstrong{{\tmem{Magic Solver}}}}.

\section{Conclusion}

So we have designed a method to solve the directed max problem. We can also
use our simple oracle to implement an improved one as showed in [E]. Since
$\rho = \sqrt{27 m / \varepsilon}$, the running time of our method is
$\tilde{O} \left( m^{4 / 3} \varepsilon^{- 3} \right)$ plus $\tilde{O} \left(
m \right)$ (for flow-cycle-canceling) equal to $\tilde{O} \left( m^{4 / 3}
\varepsilon^{- 3} \right)$.

{\noindent}\tmtextbf{Theorem. }\tmtextit{For any $0 < \varepsilon < 1 / 2$,
the directed max flow problem can be $\left( 1 - \varepsilon
\right)$-approximated in $\tilde{O} \left( m^{4 / 3} \varepsilon^{- 3}
\right)$ time.}{\hspace*{\fill}}{\medskip}

As in [E], we can combine this with the smoothing and sampling techniques of
Karger to obtain an $\tilde{O} \left( \tmop{mn}^{1 / 3} \varepsilon^{- 11 / 3}
\right)$-time algorithm.

\end{document}